\documentclass[fleqn,usenatbib]{mnras}
\usepackage{newtxtext,newtxmath}
\usepackage[T1]{fontenc}
\usepackage{ae,aecompl}
\usepackage{graphicx}
\usepackage{amsmath}	
\usepackage{amssymb}
\usepackage{array}
\usepackage{makecell}
\usepackage{rotating}
\usepackage{longtable}
\usepackage{multirow}
\usepackage{ulem}
\usepackage{xcolor}

\newcommand{\totalflips}{30}
\newcommand{\fliprightway}{27}
\newcommand{\flipsbb}{27} 
\newcommand{\flipsHIIonsky}{26} 
\newcommand{\flipsHIIonskyexceg}{24} 
\newcommand{\flipsperfect}{23} 
\newcommand{\flipsnoHII}{5}

\newcommand{\twoPi}{$^{2}\Pi_{3/2},~J = 3/2$~}

\newcommand{\Htwo}{H$_2$}

\newcommand{\HII}{H\textsc{ii}}
\newcommand{\HI}{H\textsc{i}}
\newcommand{\kms}{km\,s$^{-1}$}
\newcommand{\cmkms}{{\rm cm}$^{-2}$\,km$^{-1}$\,s}

\newcommand{\Av}{$A_\mathrm{v}({\rm ext})$}

\title[OH satellite-line `flip']{The hydroxyl satellite-line `flip' as a tracer of expanding \HII~regions}
\author[A. Petzler et al.]{
Anita Petzler,\thanks{E-mail: anita.petzler@students.mq.edu.au (AP)}
J. R. Dawson,
Mark Wardle
\\
Department of Physics and Astronomy, and Research Centre in Astronomy, Astrophysics and Astrophotonics, Macquarie University, NSW 2109, Australia\\
}
\date{Accepted XXX. Received YYY; in original form ZZZ}
\pubyear{2020}

\begin{document}
\label{firstpage}
\pagerange{\pageref{firstpage}--\pageref{lastpage}}
\maketitle

\begin{abstract}
    Observations of the four $^{2}\Pi_{3/2},~J = 3/2$~ground state transitions of the hydroxyl radical (OH) have emerged as an informative tracer of molecular gas in the Galactic ISM. We discuss an OH spectral feature known as the `flip', in which the satellite lines at 1612 and 1720\,MHz flip -- one from emission to absorption and the other the reverse -- across a closely blended double feature. We highlight 30 examples of the flip from the literature, 27 of which exhibit the same orientation with respect to velocity: the 1720\,MHz line is seen in emission at more negative velocities. These same examples are also observed toward bright background continuum, many (perhaps all) show stimulated emission, and 23 of these are coincident in on-sky position and velocity with H\textsc{ii}~radio recombination lines. To explain these remarkable correlations we propose that the 1720\,MHz stimulated emission originates in heated and compressed post-shock gas expanding away from a central H\textsc{ii}~region, which collides with cooler and more diffuse gas hosting the 1612\,MHz stimulated emission. The foreground gas dominates the spectrum due to the bright central continuum, hence the expanding post-shock gas is blue-shifted relative to the stationary pre-shock gas. We employ non-LTE excitation modelling to examine this scenario, and find that indeed FIR emission from warm dust adjacent to the H\textsc{ii}~region radiatively pumps the 1612 MHz line in the diffuse, cool gas ahead of the expanding shock front, while collisional pumping in the warm, dense shocked gas inverts the 1720 MHz line.

\end{abstract}


\section{Introduction}\label{Sec:Intro}
    The hydroxyl radical, OH, is a sensitive tracer of the physical characteristics and dynamics of the molecular interstellar medium (ISM). OH was the first molecule discovered in the ISM \citep{Weinreb1963} and though its ground rotational state hyperfine transitions are weak, many early works used it to study the distribution and properties of Galactic molecular gas \citep{Barrett1964,Robinson1967,Rogers1967,Goss1968,Heiles1968,Heiles1969,Turner1979}. However, once the technology was developed to observe the much brighter rotational transitions of carbon monoxide -- particularly the CO\,($J=1\rightarrow 0$) transition \citep[e.g.][]{Wilson1970,Solomon1972} -- the focus of OH observations moved away from its ability to trace the molecular ISM as a whole, and more towards its ability to trace specific excitation conditions \citep{Elitzur1976b,Elitzur1976c,Guibert1978,Elitzur1978}, such as those responsible for masing in each of the four ground-rotational state transitions.

Though CO remains a very effective tracer of molecular gas in the ISM, a significant amount of molecular gas in more diffuse regions is not traced well by CO \citep[e.g.][]{Reach1994,Grenier2005,PlanckCollaboration2011,Remy2018}, but is expected to be well-traced by OH \citep{Wannier1993,Liszt1996,Barriault2010,Allen2012,Allen2015,Li2018}. This has led to a resurgence of interest in the observation and interpretation of OH as a supplementary tracer of molecular gas \citep[e.g.][]{Dawson2014a,Rugel2018,Nguyen2018,Engelke2019}.

The four \twoPi~ground state transitions of OH occur at 1612.231, 1665.402, 1667.359 and 1720.530 MHz. These transitions are generally weak, and therefore are most readily observed in absorption against bright background continuum emission. OH in the molecular ISM is almost never in local thermodynamic equilibrium (LTE), so the excitation temperatures $T_{\rm ex}$ of its transitions are not a reliable indication of the kinetic temperature of the gas $T_{\rm gas}$ \citep{Elitzur1992}. Without the ability to assume LTE, the interpretation of OH emission and absorption can be quite difficult. Thus when considering the molecular ISM, many studies focus on the stronger `main' lines at 1665 and 1667 MHz as their excitation temperatures -- though often sub-thermal -- tend to not deviate from LTE as strongly as the satellite lines under a wide range of local conditions \citep[e.g.][]{Baud1980,Barriault2010,Nguyen2018,Li2018}. In such cases reasonable assumptions may be applied to the excitation of the main lines and properties of the gas can be inferred without solving for the population of the four levels within the OH ground rotational state \citep[e.g.][]{Barriault2010,Rugel2018}.

However, the complexity of OH satellite-line excitation is sometimes an advantage, as its greater sensitivity to local environmental conditions carries information that may not be captured by the main lines. For example, stimulated emission of the OH satellite lines (i.e. when $T_{\rm ex} < 0$) has been studied extensively in the context of OH masers \citep[e.g.][]{Elitzur1976b,Elitzur1976c,Guibert1978,Pavlakis1996a}. Stimulated emission in the 1612 MHz line is generally interpreted as indicating a radiatively-dominated environment, while stimulated emission in the 1720 MHz line is interpreted as indicating a collisionally-dominated environment \citep{Elitzur1992}. High-gain 1612 MHz masers (i.e. where optical depth $|\tau| \gg 1$) trace both active star formation \citep{Caswell1999} and evolved stellar sources such as OH-IR stars \citep{Elitzur1976c}. High-gain 1720 MHz masers trace the shocked gas of supernova remnants \citep{Frail1994,Frail1996}. Widespread weak satellite-line maser emission (i.e. where $|\tau| < 1$) is also observed in the Milky Way \citep[e.g.][]{Turner1982, Dawson2014a, Walsh2016, Rugel2018}.

\begin{figure}
	\centering
	\includegraphics[trim={.4cm .4cm 1cm 1.1cm}, clip=true,width=\linewidth]{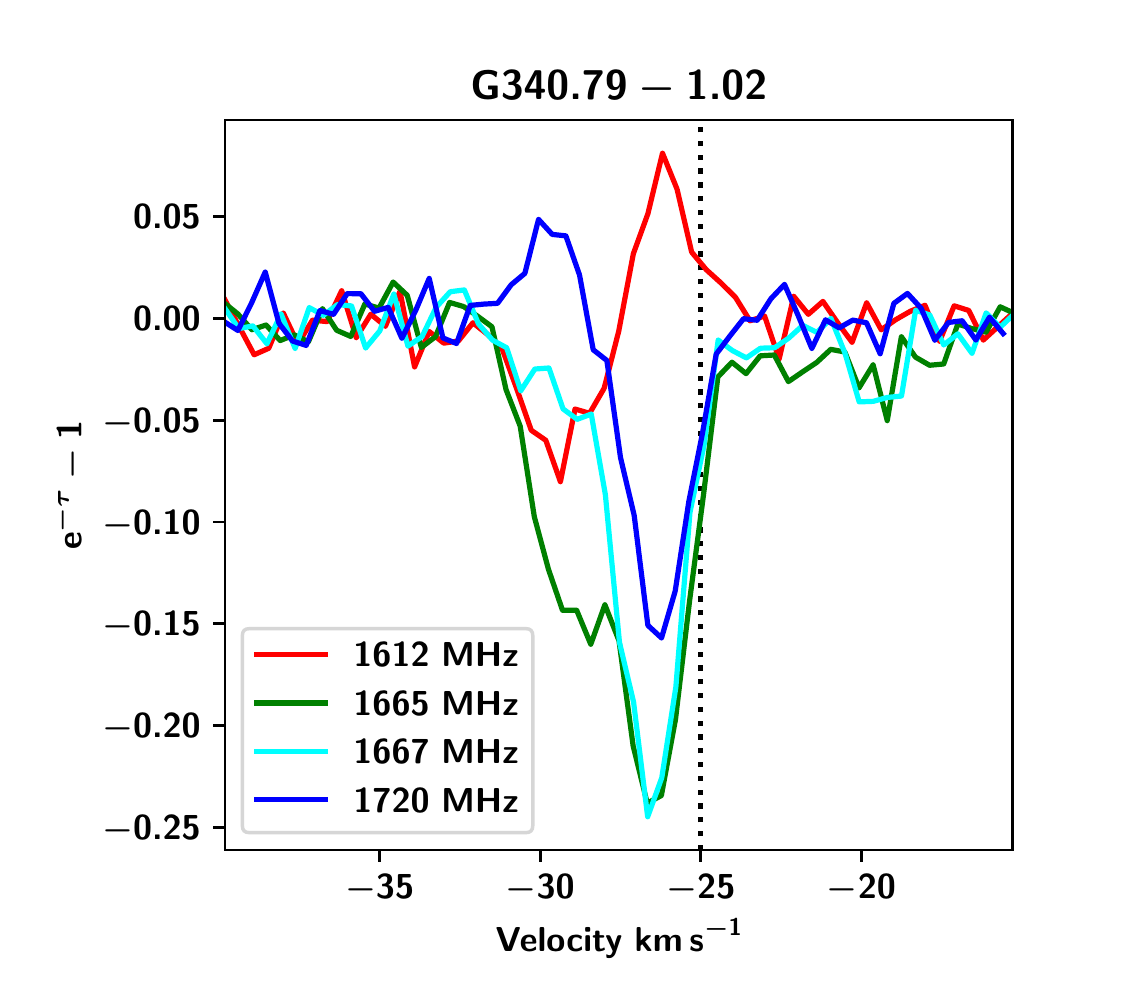}
	\caption{Optical depth observations of hydroxyl made with the Australia Telescope Compact Array towards the bright background continuum source G340.79-1.02 from Petzler et al. (2020, in prep.) as an illustrative example of the OH satellite-line flip. The 1612 MHz line can be seen in absorption at more negative velocities, transitioning to stimulated emission at -28 \kms. The 1720 MHz line shows the opposite behaviour with stimulated emission at more negative velocities, transitioning to absorption at -28 \kms. Both main lines (1665 and 1667 MHz) are in absorption but clearly indicate the presence of two blended velocity components, with peaks roughly corresponding to the peak emission in each satellite line. A radio-recombination line is detected toward this source, centred at -25 \kms~(dotted line), indicating the presence of an \HII~region.}
	\label{fig:Flip_sample}
\end{figure}

Here we present an example of a peculiar OH profile shape sometimes known as the satellite-line `flip' -- first described by \citet{Caswell1975}, with the first examples seen in the literature almost a decade earlier \citep{Goss1967}. A typical example of the profile is given in Fig. \ref{fig:Flip_sample}. On one side of the feature the 1612 MHz line is seen in absorption while the 1720 MHz line is in emission. The lines then flip their relative orientation to have the 1612 MHz line in emission and the 1720 MHz line in absorption, all within a closely blended double feature. Meanwhile, both main lines are in absorption, often obscuring the fact that two velocity components exist. 

\citet{vanLangevelde1995}~carried out non-LTE excitation calculations in the context of an extragalactic example of the flip towards the nucleus of Cen A (G200.63-42.76). They tested a range of number densities, column densities, gas temperatures and radiation fields with an assumed velocity dispersion of $\Delta V=1\,$\kms, and found that in models where a flip could be achieved, it occurred where two gas components lay on either side of $N_{\rm OH}/\Delta V \approx 10^{15}\,$\cmkms, in broad agreement with \citet{Elitzur1976b}. They therefore interpret the flip as indicating $N_{\rm OH}/\Delta V \approx 10^{15}\,$\cmkms. Other than this work and those that apply its conclusions \citep[e.g.][]{Frayer1998, Rugel2018}, no local astrophysical process has been proposed to account for the flip. In Section \ref{Sec:Model}~we outline our own non-LTE excitation modelling and find that the effects of line overlap complicate the picture reported by \citet{vanLangevelde1995}.

In this work we identify, from our own observations and from the literature, \totalflips~examples of the satellite-line flip \citep{Goss1967,Caswell1974, Caswell1975, Turner1979,vanLangevelde1995, Frayer1998, Brooks2001, Dawson2014a, Rugel2018,Ogbodo2020} presented in Table \ref{Tab:Flip}. All but two extragalactic examples (towards G200.63-42.76 \citep{vanLangevelde1995}~and towards G277.81+32.42 \citep{Frayer1998}) are within our Galaxy. With only three exceptions, all of these examples are observations of OH gas towards bright background continuum sources, most of which may be interpreted as locally associated \HII~regions. The bright background continuum suggests that the observed emission is stimulated emission (i.e. indicating a population inversion where optical depth $\tau <0$~and $T_{\rm ex}<0$), as opposed to thermal emission (i.e. where $\tau >0$~and $T_{\rm ex} > T_{\rm c}$).

We note the hitherto unnoticed fact that \textit{all} of the flips observed towards bright background continuum sources (\flipsbb/\totalflips) exhibit the same orientation in velocity: the 1720 MHz inversion is seen at more negative velocities and the 1612 MHz inversion at more positive velocities. This remarkable trend leads us to propose a physical explanation of the flip where the 1612 MHz and 1720 MHz inversions originate in pre- and post-shock gas (respectively) surrounding an \HII~region. To justify this model we first outline the key factors affecting OH excitation in the context of the satellite-line flip in Section \ref{Sec:Excitation}. We outline our physical model in detail in Section \ref{Sec:Model}, discuss its implications and limitations in Section \ref{Sec:Discussion} and present our conclusions in Section \ref{Sec:Conclusions}.

\begin{table*}
\centering
\begin{tabular}{rrcccccccccc}
\hline
$l\,^{\circ}$&$b\,^{\circ\; a}$&\multicolumn{2}{c}{$v$ (\kms)$\,^{b}$}&\multicolumn{2}{c}{1612 MHz}&\multicolumn{2}{c}{1720 MHz}&Units$^c$&$\theta_{1/2}\,^d$&\HII~Assoc.$^e$&Ref.$^{f}$\\
&&Blue&Red&Blue&Red&Blue&Red&&arcmin&Y/N\\
\hline
8.10&0.20&13&17.8&-0.15&0.11&0.25&-0.37&K&18.8&Y&A\\
12.80&-0.20&32.5&38.5&-0.2&0.4&0.64&-0.3&Jy&0.12$\dagger$&Y&B\\
14.00&-0.60&17.8&22.9&-0.25&0.13&0.14&-0.22&K&18.8&Y&A\\
19.08&-0.29&62&71&-10&10&10&-15&K$^*$&0.77$\dagger$&Y&C\\
19.61&-0.23&43.9&48.2&-0.32&0.1&0.56&-0.15&Jy&0.12$\dagger$&Y&B\\
30.50&0.00&90&95.2&-0.52&0.34&0.18&-0.34&K&18.8&Y&A\\
30.78&-0.14&90&95&-1&0.3&0.6&-0.9&K&33.4&Y&D\\
32.80&0.19&7&20&-15&15&70&-60&K$^*$&0.77$\dagger$&Y&C\\
35.20&-1.70&39.8&43.9&-0.18&0.19&0.28&-0.36&K&18.8&Y&A\\
48.92&-0.28&5.6&6.5&-0.165&0.029&0.127&-0.104&$e^{- \tau}-1$&0.5&N&E\\
49.37&-0.30&63&66&-90&100&90&-80&K$^*$&0.77$\dagger$&Y&C\\
133.70&1.20&-40.6&-38&-1&0.4&3.06&-0.44&K&18.8&Y&A\\
172.80&-13.24&5.3&6.8&0.2&-0.033&-0.067&0.067&K&3&N&F\\
173.40&-13.26&5&8&0.133&-0.04&-0.023&0.057&K&8.2&N&G\\
175.83&-9.36&7.1&7.8&0.003&-0.017&-0.043&0.031&$e^{- \tau}-1$&0.5&N&E\\
200.63&-42.76&550&555&-0.11&0.1&0.13&-0.1&Jy&0.11$\dagger$&Y&H\\
267.90&-1.10&0.1&2.1&-1.1&1&0.9&-4.9&K&12&Y&I\\
277.81&32.42&200&240&-0.015&0.01&0.015&-0.007&Jy&0.02$\dagger$&Y&J\\
316.80&0.00&-38&-35&-2.5&1.9&1.9&-1.9&K&12.6&Y&K\\
332.71&-0.71&-52&-48&-6&5&1&-9&K&2.5&Y&L\\
333.09&-0.50&-56&-51&-10&5&5&-7&K&2.5&Y&L\\
333.09&-0.50&-43&-39&-3&2&3&-3&K&2.5&Y&L\\
336.95&-0.20&-83&-75&-0.25&0.25&0.25&-0.3&K&12.6&Y&M\\
336.95&-0.20&-43&-38&-0.5&0.5&0.25&-0.75&K&12.6&N&M\\
337.10&-0.20&-44&-36.4&-0.3&0.4&0.1&-0.54&K&18.8&Y&A\\
337.80&-0.10&-53&-36.2&-0.3&0.53&0.23&-0.6&K&18.8&Y&A\\
340.79&-1.02&-29.2&-26.5&-0.07&0.081&0.038&-0.148&$e^{- \tau}-1$&0.5$\dagger$&Y&N\\
342.00&-0.20&-33&-23&-0.4&0.1&0.2&-0.2&K&12.6&Y&M\\
351.42&0.66&-6.1&-1&-0.28&0.1&0.19&-0.18&$e^{- \tau}-1$&0.37$\dagger$&Y&O\\
353.41&-0.30&-18.3&-15.2&-0.063&0.081&0.166&-0.125&$e^{- \tau}-1$&0.5$\dagger$&Y&N\\
\hline
\end{tabular}
\caption{Spectra exhibiting the satellite-line flip: where the conjugate relationship between the satellite lines flips across a closely-blended double feature. 
$^{a}$Locations of each flip are given in Galactic coordinates. 
$^{b}$The velocities of the more blue-shifted component and the more red-shifted component are given along with the peak values measured at 1612 and 1720 MHz. As these examples are drawn from disparate publications, the values of these peaks are approximate. $^c$ Units of measurement are either: continuum subtracted brightness temperature (K, $^*$converted from Jy beam$^{-1}$), continuum subtracted flux density (Jy), or optical depth ($e^{-\tau} - 1$). In all cases a positive value indicates relative emission and a negative value indicates relative absorption. 
$^{d}$The resolution of the flip observation is given in arcmin, and $\dagger$~indicates interferometric observations. 
$^e$The presence or absence of an associated \HII~region is indicated. 
$^{f}$References: 
(A) \citet{Turner1979}, 
(B) \citet{Ogbodo2020},
(C) \citet{Rugel2018}, 
(D) \citet{Goss1967}, 
(E) Petzler et al. (2020 in prep), 
(F) \citet{Xu2016}, 
(G) \citet{Ebisawa2019}, 
(H) \citet{vanLangevelde1995}, 
(I) \citet{Manchester1970}, 
(J) \citet{Frayer1998}, 
(K) \citet{Caswell1974}, 
(L) Cunningham M.R. (private communication), 
(M) \citet{Dawson2014a}, 
(N) Petzler et al. (2020 in prep), 
(O) \citet{Brooks2001}.}
	\label{Tab:Flip}
\end{table*}

\section{OH satellite-line excitation}\label{Sec:Excitation}
    In this section we outline the level population requirements for satellite-line inversion as seen in the satellite-line flip, and the de-excitation pathways that lead to these populations \citep{Elitzur1976b,Elitzur1976c, Guibert1978}. We then discuss the environmental conditions that influence the relative dominance of these de-excitation pathways, and how these could manifest in the observed satellite-line flip. An exhaustive discussion of this subject can be found in \citet{Elitzur1992}. 

\begin{figure}
	\centering
	\includegraphics[trim={4cm 10cm 3cm 14cm}, clip=true,width=\linewidth]{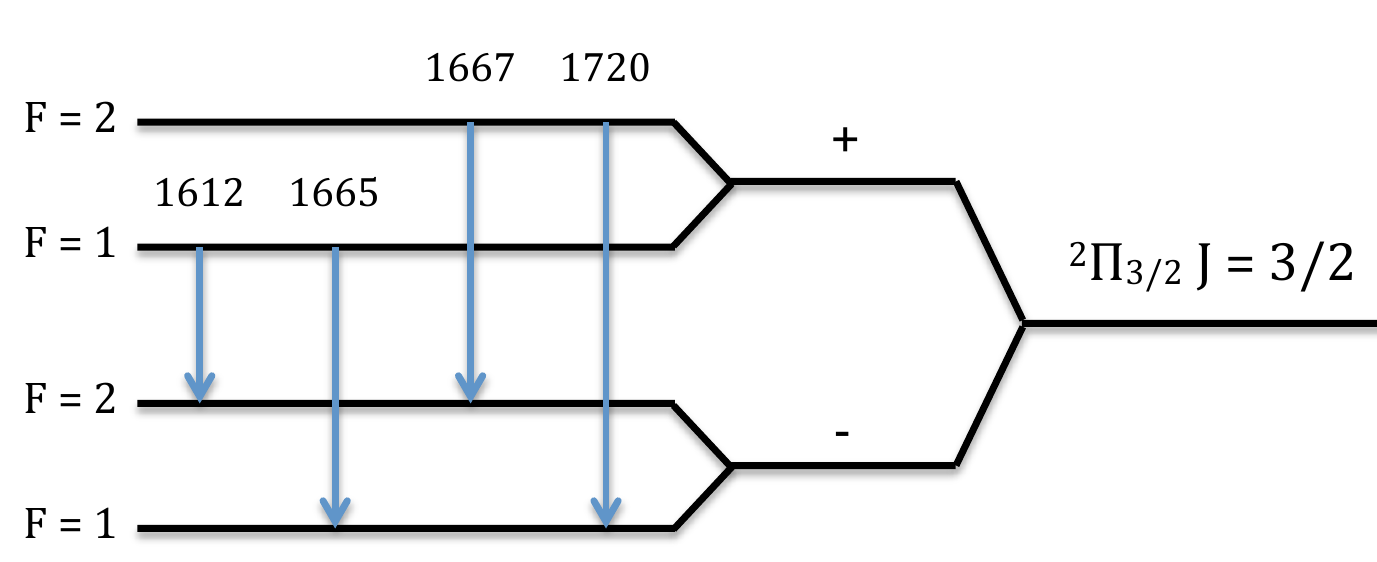}
	\caption{Energy level diagram of the \twoPi ground state of hydroxyl. The ground state is split into four levels due to $\Lambda$-doubling and hyperfine splitting, with 4 allowed transitions between these levels: `main' lines at 1665.402 and 1667.359 MHz, and `satellite' lines at 1612.231 and 1720.530~MHz. The degeneracies of the levels are given by $g = 2F+1$.}
	\label{fig:OHlevels}
\end{figure}

The OH \twoPi~ground rotational state is split into four levels via $\Lambda$-doubling and hyperfine splitting. These levels, as well as their allowed transitions at 1612, 1665, 1667 and 1720 MHz, are illustrated in Fig. \ref{fig:OHlevels}. Due to the different degeneracies of its upper and lower levels, inversion in the 1612 MHz line is achieved when the population of its upper level is greater than $3/5$~that of its lower level. Inversion of the 1720 MHz line is achieved when the population of its upper level is greater than $5/3$~that of its lower level \citep{Elitzur1992}.

\begin{figure*}
	\begin{center}
	\includegraphics[trim={1cm 9cm 1cm 8cm},clip=true,width=0.7\linewidth]{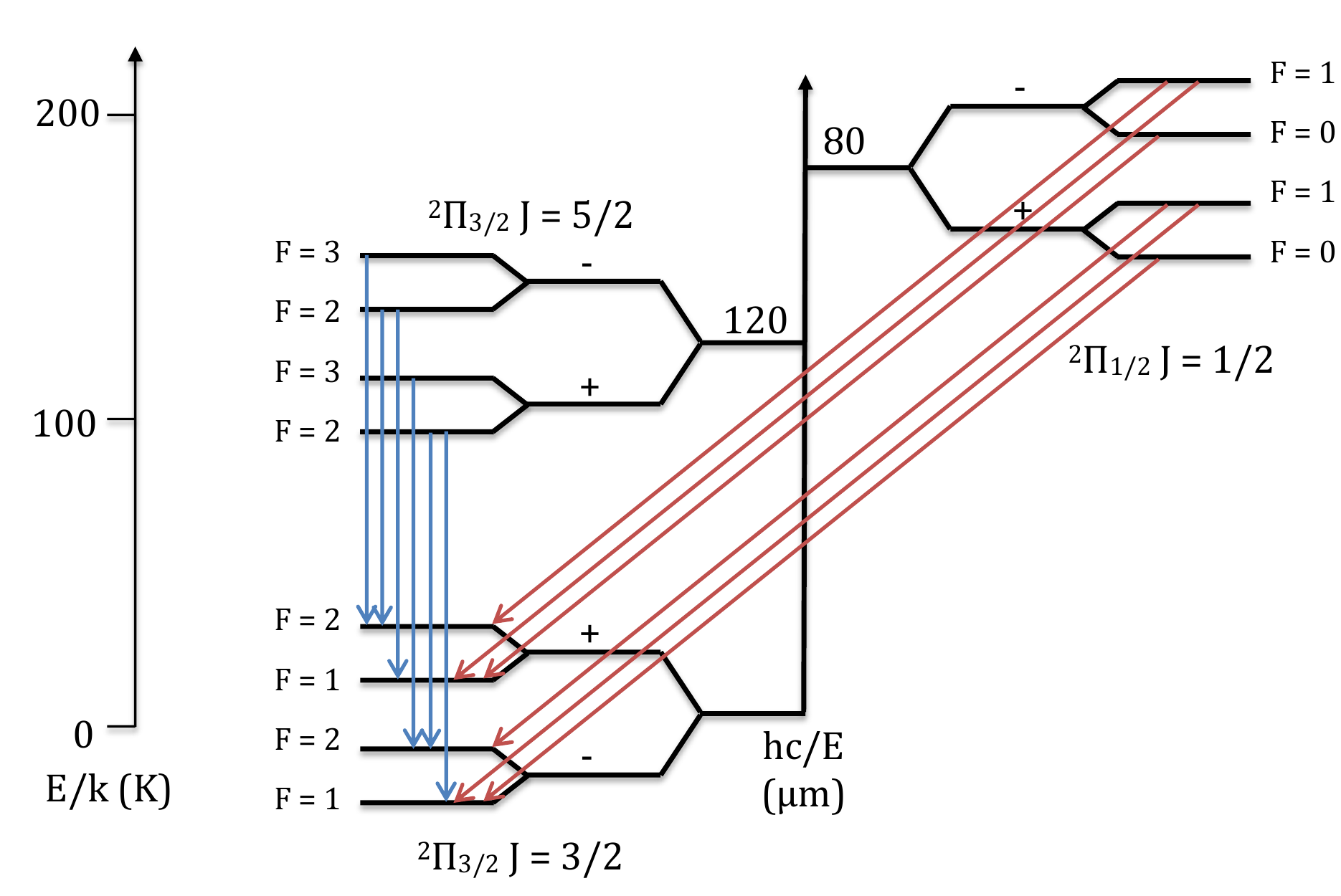}
	\end{center}
	\caption{Schematic of the three lowest rotational states of OH, as well as the $\Lambda$~and hyperfine splitting of those states. Excitations above the \twoPi ground state will cascade back down to it via the $^2\Pi_{3/2},~J=5/2$~state, or the $^2\Pi_{1/2},~J=1/2$~state. Allowable transitions are those where parity is changed and $|\Delta F|$ = 1, 0; shown in blue at left and red on the schematic. The energy scale is given at left in kelvin, and the wavelengths of the IR transitions are shown at centre in $\mu$m. The splittings of the $\Lambda$~and hyperfine levels are greatly exaggerated for clarity.}
	\label{fig:rotational_ladder}
\end{figure*}

The relative population of the OH ground state hyperfine levels in the ISM is largely determined by cascade pathways into the ground state from previously-excited higher rotational states. As all cascades into the ground state will pass through the first excited $^2\Pi_{3/2},~J=5/2$~or the second excited $^2\Pi_{1/2},~J=1/2$~rotational states \citep{Elitzur1992}, only these are illustrated in Fig. \ref{fig:rotational_ladder}. Also illustrated in Fig. \ref{fig:rotational_ladder}~are the allowed transitions (determined by quantum mechanical selection rules) from the levels within those excited rotational states into the levels of the ground state. If the radiative decay pathways from the first and second excited rotational states into the ground state are optically thick\footnote{In this context a transition between rotational states is `optically thick' if a photon emitted by an OH molecule in the upper rotational state has a high probability of being reabsorbed before leaving the cloud: this is determined by the population of the lower level, i.e. the ground rotational state.}, then the rates of individual transitions within each pathway are independent of line strength due to photon trapping, and depend only on the number of possible transitions into the ground state \citep{Elitzur1976b}. 

As illustrated in Fig. \ref{fig:OHlevels}, the main-line transitions at 1665 and 1667 MHz occur between upper and lower levels with the same total angular momentum quantum number $F$, while the satellite-line transitions occur between levels with different $F$~number. The $F$~numbers of the different levels of the ground and excited rotational states (see Fig. \ref{fig:rotational_ladder}) introduce an asymmetry in the possible transitions into the ground state for the satellite lines but not for the main lines: there are more allowed transitions from the first excited $^2\Pi_{3/2},~J=5/2$~state into the upper level of the 1720 MHz transition than there are into its lower level. Similarly, there are more allowed transitions from the second excited $^2\Pi_{1/2},~J=1/2$~state into the upper level of the 1612 MHz transition than there are into its lower level. Environmental factors that preference the cascade pathway from one of the excited rotational states over the other can therefore provide a pumping mechanism that inverts the 1720 or 1612 MHz transitions. In general, when one of the satellite-lines is inverted in this manner, the other satellite-line is sub-thermally excited, leading to the `conjugate' behaviour often observed in the satellite lines \citep[e.g.][]{vanLangevelde1995, Ebisawa2019}, the `flip' being one manifestation of this behaviour.

There are three main physical mechanisms by which one of these pathways is preferred and hence one of the satellite lines can be inverted:
\begin{enumerate}

\item In radiatively-dominated environments where a wide range of photon energies are available (i.e. far-infrared emission from warm dust), and where the transitions into the ground state from both the first and second excited rotational levels are optically thick, the two cascade pathways are equally likely, and the populations of each of the levels of the ground state will be roughly equal. Due to the level degeneracies (3 and 5 for the upper and lower levels of the 1612 MHz transition, respectively, and the reverse for the 1720 MHz transition) this leads to sub-thermal excitation of the 1720 MHz line and inversion of the 1612 MHz line \citep{Elitzur1976c}.

\item In collisionally-dominated environments where the transitions into the ground state from both the first and second excited rotational levels are optically thick, high temperatures can invert the 1612 MHz line and lower temperatures can invert the 1720 MHz line \citep{Elitzur1976b}. Lower temperatures lead to a sufficient reduction in excitation into higher rotational states that the pathway via the $^2\Pi_{3/2},~J=5/2$~state is left to dominate the population of the ground state levels. This pathway will in effect transfer molecules from the $F=1$~levels of the ground state to the $F=2$~levels, leading to sub-thermal excitation in the 1612 MHz line and inversion of the 1720 MHz line.

\item In cases where the 1612 MHz pumping mechanism is active, it can be deactivated by lowering the OH column, which can then allow the 1720 MHz line to invert. The line strengths for transitions to the ground state from the second excited rotational state are about an order of magnitude lower than those from the first excited rotational state. As a consequence, for a given velocity dispersion there exists a range of column densities for which transitions from the second excited ($^{2}\Pi_{1/2},~J = 1/2$) rotational level are optically thin while transitions from the first excited ($^{2}\Pi_{3/2},~J = 5/2$) level are optically thick. \citet{vanLangevelde1995} found that this range is $N_{\rm OH}/\Delta V \approx 1$--$9 \times 10^{14}$\,\cmkms. In this case, the rate of transitions from the second excited state to the ground state will be determined by their individual line strengths, which will tend to maintain the original population of the ground state levels. This prevents the transfer of molecules from the $F=2$~levels of the ground state to the $F=1$~levels, thus disabling the 1612 MHz pumping mechanism \citep{Elitzur1976b}. 
\end{enumerate}

In the context of the flip towards the nucleus of Cen A (G200.63-42.76), \citet{vanLangevelde1995} performed non-LTE calculations of OH assuming $\Delta V = 1$\,\kms. Their models included a radiation field characterised by two IR components: one with dust temperature $T_{\rm dust}=43\,$K and external visual extinction \Av~$=30\,$mag, and the other with $T_{\rm dust}=150\,$K and \Av~$=3\,$mag. This choice was motivated by the excess 25 and 12\,${\rm \mu m}$~emission observed along the disk of Cen A by \citet{Marston1988}. These mid-IR components allowed for excitation into the $^2 \Pi _{1/2}~J=5/2$~rotational state -- the highest state included in their modelling -- through absorption of the 35\,${\rm \mu m}$~line. These OH molecules then cascade down to the ground rotational state via the second excited rotational state, allowing the inversion of the 1612 MHz line. They found that this 1612 MHz inversion could then be disabled and the 1720 MHz line inverted when $N_{\rm OH}/\Delta V \approx 10^{15}\,$\cmkms. 

As we discuss in the following section, we also find that the flip can be produced by changes in $N_{\rm OH}/\Delta V$~in radiatively dominated environments, but due to line overlap this trend is only seen while $\Delta V \lesssim 1$\,\kms. Instead we find that a flip can also be generated by a change in a combination of parameters such as gas temperature and number density, and that these changes can naturally exist in the environment around an expanding \HII~region.

\section{Physical explanation of the flip}\label{Sec:Model}
\begin{figure}
	\centering
	\includegraphics[trim={1.8cm 0.6cm 2cm 2cm}, clip=true,width = \linewidth]{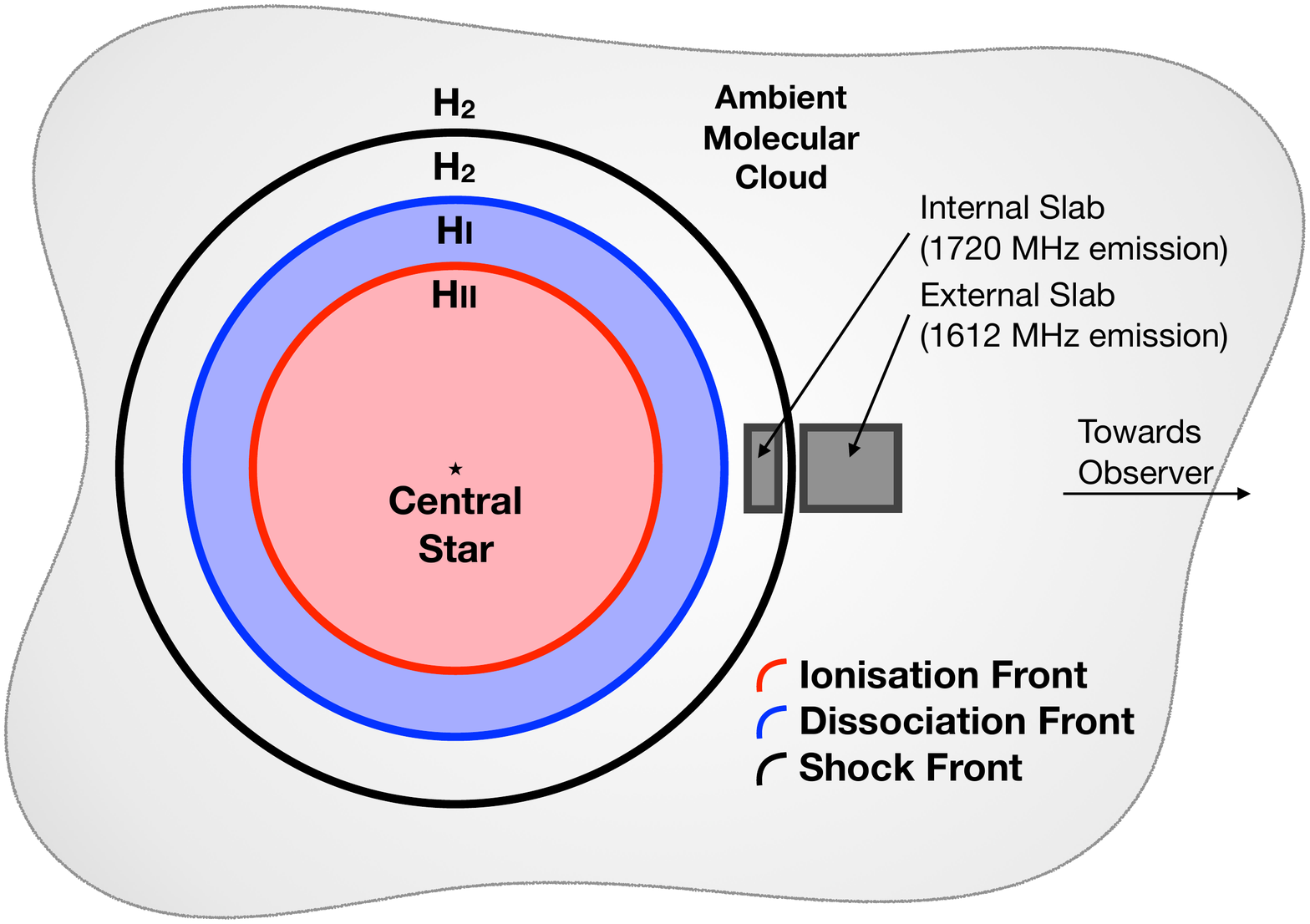}
	\caption{Cartoon of an expanding \HII~region and associated fronts.
	The relative positions of the ionisation front (red), the dissociation front (blue) and the shock front (black) expanding from the central star into an ambient molecular medium are shown. The locations of parallel slabs internal (post-shock) and external (pre-shock) to the shock front modelled in Fig. \ref{fig:contours} are indicated. The widths of the \HI~and post-shock \Htwo~regions are exaggerated for clarity.}
	\label{fig:flip_cartoon}
\end{figure}

    We propose that the satellite-line flip is best explained by a physical model where two parcels of gas exist on either side of a shock expanding away from a central \HII~region, as illustrated by the cartoon in Fig. \ref{fig:flip_cartoon}. The \HII~region provides the radiation required to invert the 1612 MHz line in the surrounding ambient molecular cloud, while the increased gas temperature and number density in the post-shock gas disable the 1612 MHz inversion and invert the 1720 MHz line. The central \HII~region then provides the bright background continuum which leads the foreground emission and absorption to dominate the observed brightness temperature, thus accounting for the striking velocity bias seen in examples of the satellite-line flip. Our reasoning and evidence are detailed in this section.

The flip is most commonly observed with the 1720 MHz line inverted at more negative velocities and the 1612 MHz line inverted at more positive velocities -- 90\% (\fliprightway/\totalflips) of the examples shown in Table \ref{Tab:Flip}~show this orientation. The separation in velocity suggests that the satellite line flip may be associated with a coherent velocity field, such as expansion and/or infall. The observation that most of the flips have the 1720 MHz inversion at more negative velocities implies a bias in the orientation of this velocity field with respect to the observer. As examples of the flip are observed in all quadrants of the Milky Way, this bias is not due to the large-scale Galactic velocity field. Instead, we note that all of the flips that demonstrate this velocity bias are also observed against bright background continuum (\flipsbb/\totalflips). When the continuum brightness temperature $T_{\rm c} \gg |T_{\rm ex}|$, absorption or stimulated emission from the foreground gas will dominate the observed line brightness temperature. Therefore, if the source of background continuum was also the source of the velocity field, this could provide a natural explanation for the observed bias in the orientation of the flip -- in principle, a physically-associated source of continuum (e.g. an embedded \HII~region) could provide the necessary bright background continuum and drive expansion. 

This led us to consider whether there was evidence for a local association between the OH gas and the background continuum illuminating it: i.e. whether the continuum source could be identified as an \HII~region and localised in on-sky position and velocity to the observed flip.

\begin{figure}
    \centering
    \includegraphics[trim = {1.5cm 0.7cm 1cm 2cm}, clip = true,width = \linewidth]{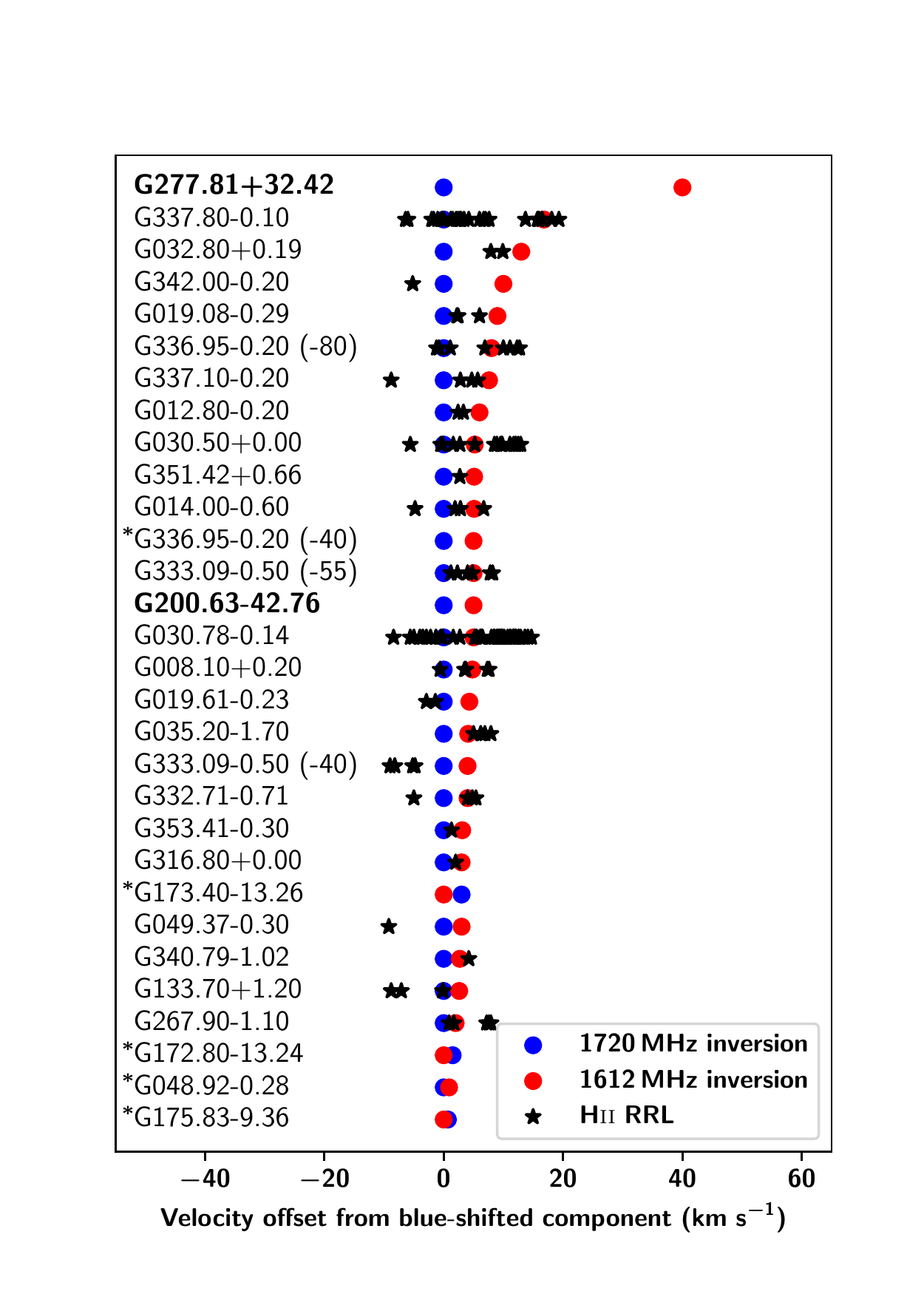}
    \caption{Velocity separation of the 1720 MHz emission (blue circles) and 1612 MHz emission (red circles) components of the examples of the satellite-line flip given in Table \ref{Tab:Flip}. The velocities of RRL for \HII~regions overlapping in on-sky position with each example of the flip are indicated by black stars. Extragalactic examples of the flip are highlighted with boldface type, and those without associated \HII~RRL are indicated with an asterisk ($^{*}$). Details of how these associations were determined are given in the text. All velocities are plotted as offsets from the blue-shifted component of the flip. The flips are plotted in descending order of the velocity offset of their red-shifted component to illustrate the distribution of velocity separation. All \HII~data were obtained from the Wide-field Infrared Survey Explorer \citep[WISE][]{Anderson2014,Anderson2015,Anderson2018b}~and from the Southern \HII~Region Discovery Survey \citep[SHRDS][]{Wenger2019}.}
    \label{fig:HII}
\end{figure}

All but one of the examples of the flip towards bright background continuum (\flipsHIIonsky/\flipsbb) coincide positionally on the sky with the location of a known \HII~region. A flip was considered to `coincide' with an \HII~region if the half-power beam-width with which it was observed overlapped with the recorded radius of an \HII~region. In the case of the two extragalactic examples, specific \HII~regions are not resolved but both are associated with large molecular structures that contain \HII~regions \citep{Hodge1983,Ulvestad1991}. Excluding these two extragalactic examples, all but one of the flips associated on the sky with specific \HII~regions (\flipsperfect/\flipsHIIonskyexceg) also have recorded radio recombination lines (RRL) within $\pm 10$\,\kms~of one or other component of the flip. The relative velocities of the components of the flips with the RRL of on-sky overlapping \HII~regions are illustrated in Fig. \ref{fig:HII}. All \HII~data were obtained from the Wide-field Infrared Survey Explorer \citep[WISE][]{Anderson2014,Anderson2015,Anderson2018b}~and from the Southern \HII~Region Discovery Survey \citep[SHRDS][]{Wenger2019}. Any more precise association with individual \HII~regions was not possible given the large full width at half-maximum of radio recombination lines (WISE median $\Delta V = 25$\,\kms) compared to the average separation of the flips, and velocity blending along the line of sight.

We then undertook non-LTE molecular excitation modelling to determine if the conditions expected in the molecular gas surrounding an \HII~region could give rise to the flip. Our modelling explored `reasonable' values of gas temperature ($T_{\rm gas}=5$--$150$\,K), number density ($n_{\rm H_2}=1$--$10^6\,{\rm cm}^{-3}$), velocity dispersion ($\Delta V=0.5$--$10$\,\kms), column density ($N_{\rm OH}=10^{12}$--$10^{18}\,{\rm cm}^{-2}$), internal and external dust temperatures ($T_{\rm dust}\textrm{(int)}=10$--$50\,$K and $T_{\rm dust}\textrm{(ext)}=10$--$100\,$K, respectively) and external visual extinction (\Av~$=0.1$--$10\,$mag). The range of parameters tested were chosen to be consistent with expected conditions in the molecular ISM and the previous findings of \citet{Elitzur1976c}~and \citet{vanLangevelde1995}. It was also informed by 1D hydrodynamical modelling of the shock, dissociation and ionisation fronts surrounding an expanding \HII~region \citep{Hosokawa2005,Hosokawa2006}, as well as modelling and observations of dust temperatures around \HII~regions \citep{Churchwell1990,Okumura1996,Anderson2012,Deharveng2012}. 

\citet{Hosokawa2006} modelled the expansion of an \HII~region surrounding solitary O-type stars in a typical host molecular cloud with $n_{\rm H_2} = 10^3\,{\rm cm}^{-3}$. Their models showed the velocity discontinuity at the shock to be $\approx 1-8$\,\kms, consistent with a large portion of the flips identified in Table~\ref{Tab:Flip}. They also found that the post-shock gas could be heated to a gas temperature 50--100\,K warmer than the pre-shock gas, and that the number density of the post-shock gas was increased by a factor of 10. 

Observations consistently find that dust temperatures in molecular clouds are typically 10--30\,K \citep{Okumura1996,Anderson2012,Deharveng2012}, and only rarely exceed 40\,K \citep{Dupac2003,Salgado2016}. We therefore considered internal dust temperatures $T_{\rm dust}(\rm{int})=10$--$50\,$K. Modelling predicts, however, that the hottest dust surrounding an \HII~region (i.e. within the photodissociation zone) could be up to several $10^2\,$K -- but only around the youngest stars and in a very thin layer \citep{Wolfire1994}. We therefore considered external dust temperatures $T_{\rm dust}(\rm{ext})= 10$--$100$\,K and visual extinction \Av~$= 0.1$--$10$\,mag, but models with high $T_{\rm dust}(\rm{ext})$~were paired with low \Av~and vice versa.

The full parameter ranges tested are outlined in Table \ref{Tab:Molex_models}, and extend from quiescent, diffuse molecular gas to shock-compressed gas nearing dissociation in close proximity to an \HII~region. Values outside these ranges that were also tested are referred to explicitly in Section \ref{Sec:Discussion}. 

\setlength{\tabcolsep}{2pt}
\begin{table}
    \centering
    \begin{tabular}{lrl}
	\hline
	property&\multicolumn{2}{c}{values explored}\\
	\hline
	$N_{\rm OH}$&$10^{12}$&$-~10^{18}\,{\rm cm^{-2}}$\\
	$\Delta V$&$0.5$&$-~10\,$\kms\\
	$T_{\rm gas}$&$5$&$-~150\,{\rm K}$\\
	$n_{\rm H_2}$&$10^{0}$&$-~10^6\,{\rm cm^{-3}}$\\
	$T_{\rm dust}$\,(internal) &$10$&$-~50\,{\rm K}$\\
	$T_{\rm dust}$\,(external) &$10$&$-~100\,{\rm K}$\\
	$A_{\rm v}$\,(external)&$0.1$&$-~10\,{\rm mag}$\\ 
	\hline
    \end{tabular}
    \caption{Properties of models explored using our molecular excitation code to approximate slabs of molecular gas surrounding an \HII~region as illustrated in Fig. \ref{fig:flip_cartoon}. The effects of gas temperature, velocity dispersion, OH column density, \Htwo~number density, internal dust temperature and radiation field as defined by external dust temperature and visual extinction on the satellite-line flip were explored across the region of parameter space indicated by the ranges given. Values of ortho : total \Htwo~ratio (0.75), OH and He abundances relative to \Htwo~($10^{-7}$~and $10^{-4}$, respectively) were kept constant. The resulting optical depths at 1612 and 1720 MHz found from this modelling are illustrated in Fig. \ref{fig:contours}.}
    \label{Tab:Molex_models}
\end{table}

Our non-LTE molecular excitation code was used to determine the level populations of the lowest 8 rotational states of OH and their hyperfine levels (32 levels in all). The code employed an escape probability approach to radiative transfer in a uniform slab including line overlap, following the treatment of \citet{Lockett1989}. Line overlap occurs when the Doppler shift due to the velocity dispersion in the gas spans the difference in frequency between the 3 or 4 IR transitions linking the same pair of rungs of the OH rotational ladder. The most relevant IR transitions are those between the first two excited rotational states and the ground rotational state (i.e. those illustrated in Fig. \ref{fig:rotational_ladder}). Line overlap disrupts the population inversions caused by the pumping mechanisms outlined in Section \ref{Sec:Excitation} by allowing photons emitted in one IR transition to be reabsorbed in a neighbouring IR transition when the linewidths exceed $\sim 1$\,\kms. The escape probability approach makes the approximation that the OH level population does not vary with depth within the slab.  The level population is solved iteratively for consistency with the rates of collisional excitation and de-excitation, spontaneous emission, and of absorption and stimulated emission implied by the mean intensity of the radiation field within the slab. The mean intensity (frequency-dependent but averaged over solid angle and position) includes the source function and absorption coefficient implied by the level population, as well as thermal emission from external and internal dust.  

\begin{figure*}
    \centering
    \begin{tabular}{ccc}
    \thead[cc]{`Low' Radiation:\\
    $T_{\rm dust}({\rm int})=20$\,K\\
    $T_{\rm dust}({\rm ext})=20$\,K\\
    \Av~$=0.3$\,mag\\
    \\
    \includegraphics[trim = {5.7cm 1.4cm 1.3cm 8.9cm}, clip = true, width=0.15\linewidth]{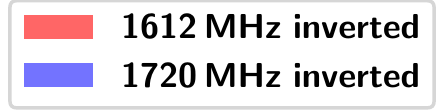}}&
    \thead[cc]{\includegraphics[trim = {-.6cm 0cm -.6cm .5cm}, clip = true, width=0.42\linewidth]{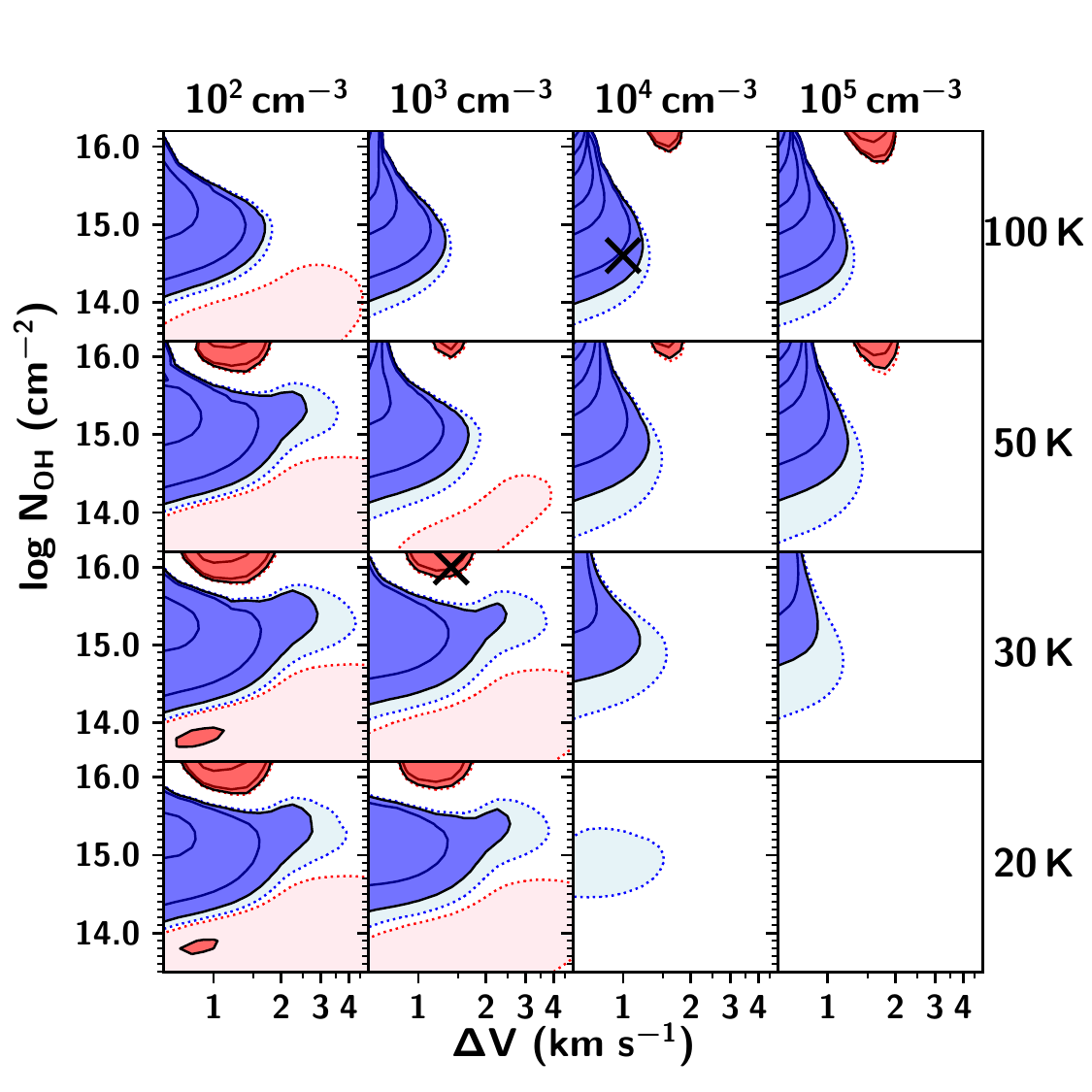}}&
    \thead[cc]{\includegraphics[trim={0.5cm 0.5cm .9cm .4cm}, clip=true, width=0.39\linewidth]{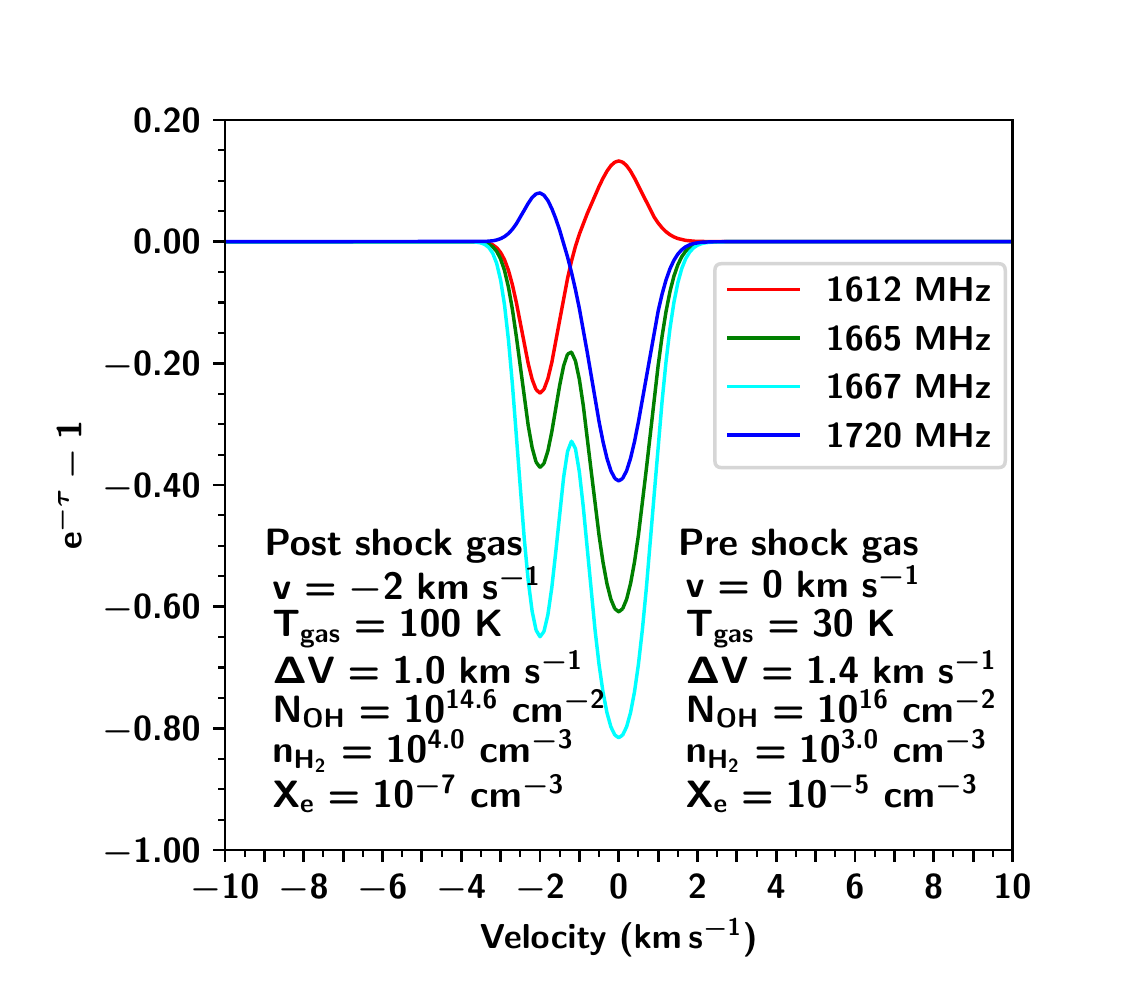}}\\
    \thead[cc]{`High' Radiation:\\
    $T_{\rm dust}({\rm int})=20$\,K\\
    $T_{\rm dust}({\rm ext})=70$\,K\\
    \Av~$=0.1$\,mag\\
    \\
    \includegraphics[trim = {5.7cm 1.4cm 1.3cm 8.9cm}, clip = true, width=0.15\linewidth]{Contour_plot_legend.pdf}}&
    \thead[cc]{\includegraphics[trim = {-.6cm 0cm -.6cm .5cm}, clip = true, width=0.42\linewidth]{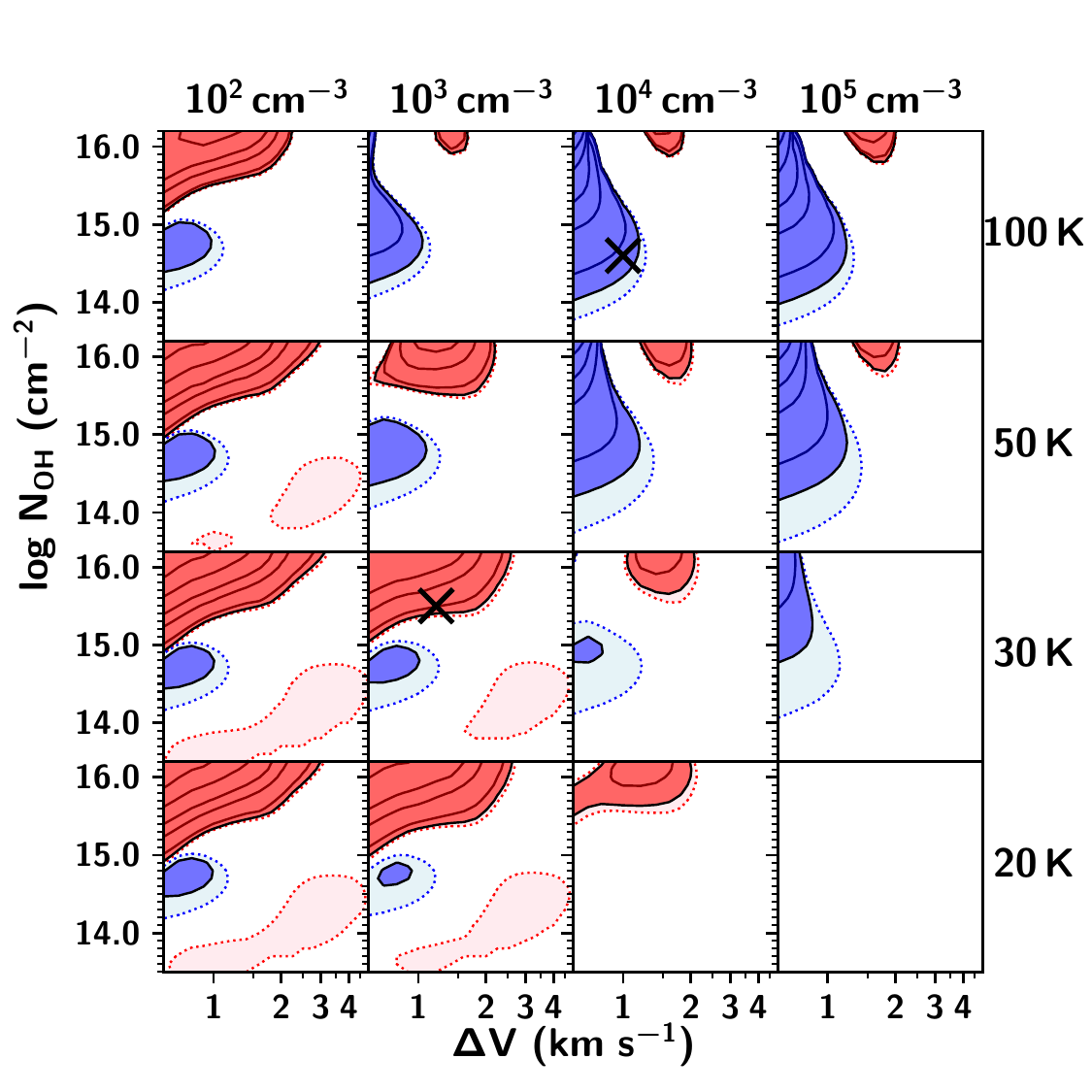}}&
    \thead[cc]{\includegraphics[trim={0.5cm 0.5cm .9cm .4cm}, clip=true, width=0.39\linewidth]{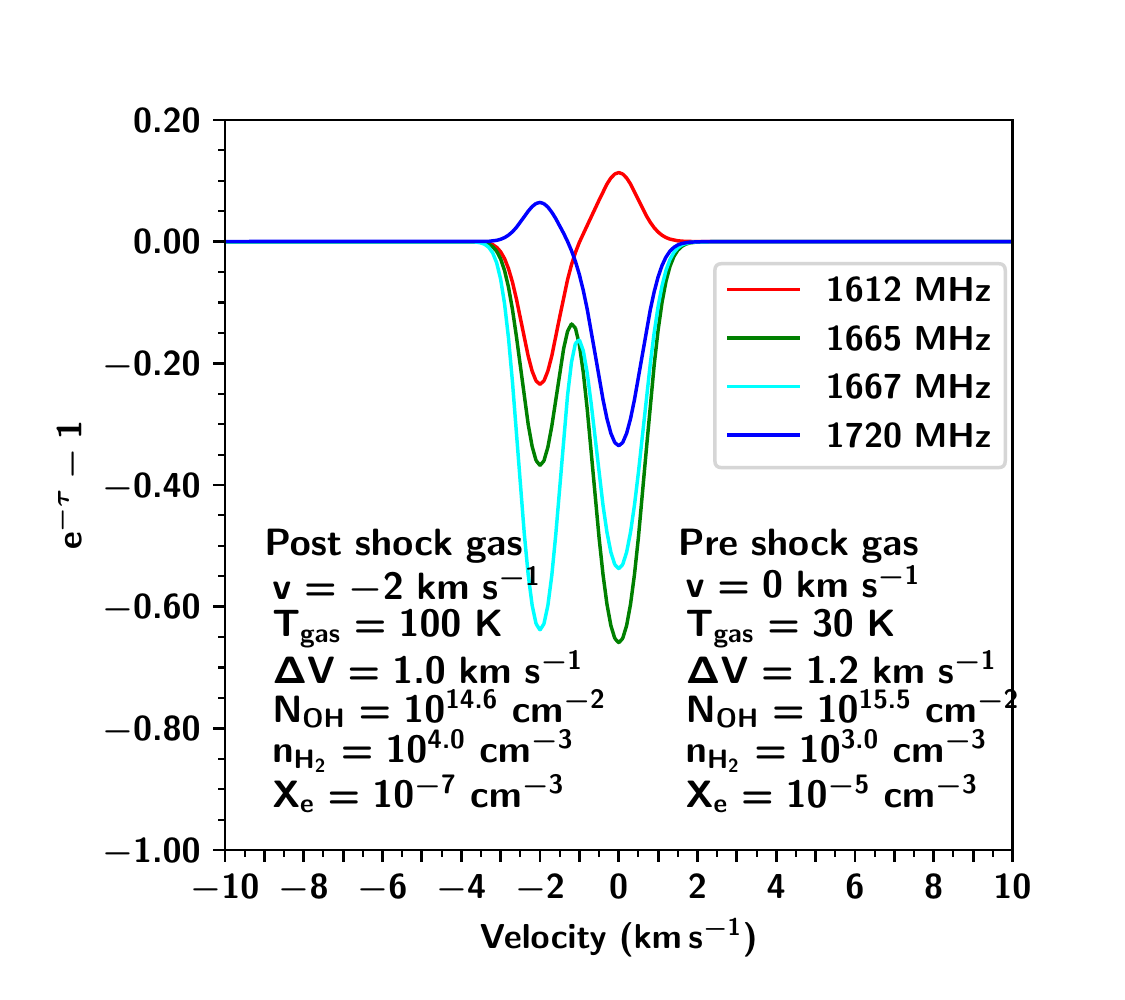}}\\
    
    \end{tabular}
    \caption{\textit{Left panels:} Regions of ${\rm log}\,N_{\rm OH}$ vs $\Delta V$~parameter space where the 1612 MHz (pink) and 1720 MHz (blue) OH satellite lines are inverted. Dotted contours and lighter shading indicate where $-0.02\leq\tau \leq 0$. Solid contours and darker shading represent areas of `significant' inversion where $\tau = -(0.02,0.1,0.5,1,2,3,4)$. The grids both have constant internal dust temperature $T_{\rm dust}(\rm int)=20$\,K. 
    From top to bottom the grids increase in external dust temperature $T_{\rm dust}(\rm ext)=20-70$\,K and decrease in visual extinction (equivalent to the thickness of the external dust layer) \Av~$=0.3-0.1$\,mag. The grids therefore represent a transition from `low' radiation at the top to `high' radiation at the bottom. These parameters were chosen to represent increasing proximity to the central ionising sources and/or differences in the properties of those sources and their surrounding ISM. The panels in each row of the grids correspond to gas temperature $T_{\rm gas} = 20,~30,~50,~100\,{\rm K}$~from bottom to top; the columns to number density $n_{\rm H_2} = 10^2,~10^3,~10^4,~10^5\,{\rm cm}^{-3}$~from left to right. The satellite-line flip can be achieved by a two component feature with one component in a pink shaded region and the other component in a blue shaded region. \textit{Right panels:} Synthetic profiles with parameters corresponding to the positions in parameter space marked by black crosses in the adjacent grid. All parameters used are given on the respective plots.}
    \label{fig:contours}
\end{figure*}

The energy levels and Einstein $A$~coefficients for the hyperfine transitions within rotational states were taken from \citet{Destombes1977}, while the infrared transitions and $A$~coefficients were taken from \cite{Brown1982}. Collisional rate coefficients with \Htwo\ were taken from \citet{Offer1994} for transitions between the lowest 6 rotational levels, with hard sphere collision rates used for transitions involving the upper two rotational levels. Collisional rate coefficients with helium were taken from \citet{Klos2007} and were computed for electrons using the prescription from \citet{Chu1976}. The infrared radiation fields from dust lying outside the slab and within the slab were represented by grey body models using the MRN dust extinction curve of \citet{Draine1984}. The input parameters were gas temperature $T_{\rm gas}$, velocity FWHM $\Delta V$ of the Gaussian line profile (which implicitly includes the contributions of thermal and ``microturbulent'' broadening), fraction of ortho-\Htwo, abundances of OH, He and electrons relative to \Htwo~($X_{\rm OH}$, $X_{\rm He}$, $X_{\rm e}$), external dust temperature $T_{\rm dust}({\rm ext})$ and external visual extinction \Av, internal dust temperature $T_{\rm dust}({\rm int})$, number density $n_{\rm H_2}$~and column density $N_{\rm OH}$. The extinction of the dust internal to the slab is computed using $A_\mathrm{v}\mathrm{(int)} = N_{\rm OH}/ X_{\rm OH} / ( 10^{21} \mathrm{cm^{-2}})$. The electron abundance relative to \Htwo~($X_{\rm e}$) declined with increasing $n_{\rm H_2}$, ranging from $X_{\rm e}=10^{-4}$~when $n_{\rm H_2}=10^2\,{\rm cm^{-3}}$~to $X_{\rm e}=10^{-7.5}$~when $n_{\rm H_2}=10^5\,{\rm cm^{-3}}$, reflecting the relative dominance of photo- and cosmic ray ionisation expected in these regimes. The fraction of ortho-\Htwo~(0.75), $X_{\rm OH}$~($n_{\rm OH}/n_{\rm H_2}=10^{-7}$) and $X_{\rm He}$~($n_{\rm He}/n_{\rm H_2}=0.2$)~were held constant for all models. Some key results of our modelling are illustrated in Fig. \ref{fig:contours}.

The grids on the left in Fig. \ref{fig:contours}~show the pattern of satellite-line inversion across the parts of parameter space most relevant to our discussion: pink regions indicate inversion of the 1612 MHz line and blue regions indicate inversion of the 1720 MHz line. The grids show two radiation regimes characterised by their internal and external dust temperatures, and external dust visual extinction. In the `low' radiation regime shown at top in Fig. \ref{fig:contours}, the dust within the modelled slab has a temperature of 20\,K, and is surrounded by a layer of external dust with a temperature of 20\,K and visual extinction \Av~$ = 0.3$\,mag. In the `high' radiation regime shown at bottom in Fig. \ref{fig:contours}, the dust within the slab also has a temperature of 20\,K but is exposed to a hotter, thinner layer of dust, with temperature 70\,K and visual extinction 0.1\,mag. These two sets of model runs are intended to capture a realistic range of conditions, potentially corresponding to increasing proximity to a central ionising source and/or differences in the properties of that source and the surrounding ISM. When we modelled the `high' radiation regime with a thinner and hotter layer of external dust we found there was little change to the pattern of satellite-line inversion.

Each of the panels within the grids illustrate the pattern of satellite-line inversion for a range of velocity dispersion ($\Delta V = 0.5$--$5$\,\kms) and column density ($N_{\rm OH}=10^{13.5}$--$10^{16.2}\,{\rm cm}^{-2}$), and how those patterns change across the number densities ($n_{\rm H_2}=10^2$--$10^5\,{\rm cm}^{-3}$) and gas temperatures ($T_{\rm gas}=
20$--$100$\,K) that might be expected in the molecular gas surrounding an \HII~region. The presence of 1612 MHz (pink) or 1720 MHz (blue) inversion in most panels of the two grids implies that the flip could potentially be produced in either radiation regime -- provided that one parcel of gas occupies a pink region and the other occupies a blue. This is demonstrated by the synthetic profiles on the right of Fig. \ref{fig:contours}~which are both broadly consistent with the examples of the flip seen in the literature. These profiles were generated assuming a 2\,\kms~velocity discontinuity between the 1612 MHz-inverted and 1720 MHz-inverted parcels of gas in the arrangement shown in Fig. \ref{fig:flip_cartoon}. The parameters of the gas are shown on the plots and indicated by black crosses on the adjacent grid. We assume that both parcels of gas responsible for the flip would be exposed to the same radiation field. 

With reference to the satellite-line pumping mechanisms outlined in Section \ref{Sec:Excitation}, the radiatively dominated regime is best illustrated at low number density ($n_{\rm H_2}\lesssim 10^4\,{\rm cm}^{-3}$) and gas temperature ($T_{\rm gas}\lesssim 50\,$K) in the bottom grid of Fig. \ref{fig:contours}. The excitation pattern is characterised by 1612 MHz inversion above a threshold value of $N_{\rm OH}/\Delta V \approx 10^{15}\,$\cmkms. Below this threshold the 1720 MHz line is able to invert, but only for a narrow range of column densities ($N_{\rm OH} \sim 10^{14}$--$10^{15}\,{\rm cm}^{-2}$) and at low velocity dispersion ($\Delta V \lesssim 1\,$\kms). This is consistent with \citet{Elitzur1976c}~and \citet{vanLangevelde1995}~only at low velocity dispersion, as neither of those works considered the effects of line overlap which disrupts the $N_{\rm OH}/\Delta V$~trend when $\Delta V \gtrsim 1\,$\kms.

The collisionally dominated regime is best illustrated at high number density ($n_{\rm H_2}\gtrsim 10^3\,{\rm cm}^{-3}$) and gas temperature ($T_{\rm gas}\gtrsim 30\,$K) in both grids of Fig. \ref{fig:contours}. The excitation pattern is characterised by 1720 MHz inversion across a wide range of column densities at velocity dispersions lower than $\sim 2\,$\kms. At higher velocity dispersions the 1720 MHz pumping mechanism is disabled by line overlap. The 1612 MHz line is strongly inverted only at high column densities ($N_{\rm OH} \sim 10^{16}\,{\rm cm}^{-2}$).

\begin{figure}
    \centering
    \includegraphics[trim={.5cm .7cm .8cm 1cm}, clip=true,width=.7\linewidth]{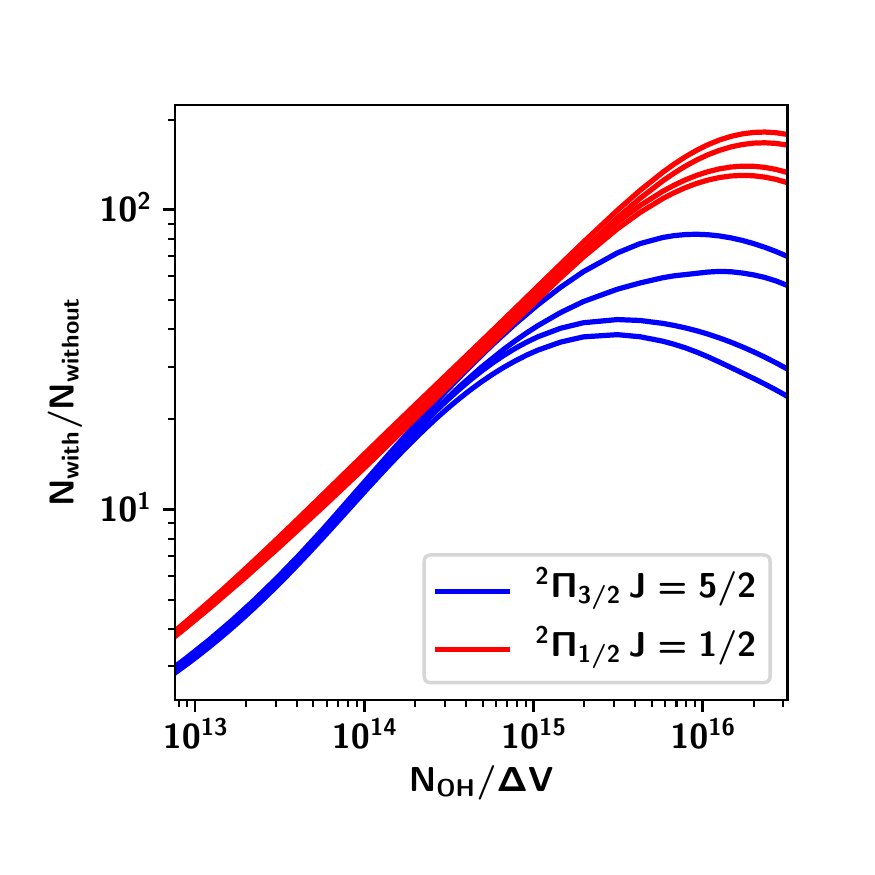}
    \caption{The ratio of level populations $N$~with the effects of internal dust and without as a function of $N_{\rm OH}/ \Delta V$~for the hyperfine levels in the first excited $^2\Pi_{3/2}\,J=5/2$~rotational state (blue) and the second excited $^2\Pi_{1/2}\,J=1/2$~rotational state (red). The values shown represent a model cloud in the `low' radiation regime (see Fig. \ref{fig:contours}) with gas kinetic temperature 30\,K, number density $10^3\,{\rm cm}^{-3}$~and velocity dispersion 1.4\,\kms.}
    \label{fig:level_ratio}
\end{figure}

In both radiation regimes there is a region of weak 1612 MHz inversion below $N_{\rm OH}/ \Delta V \approx 2 \times 10^{14}$\,\cmkms. This weak inversion is only seen when the effect of radiation from internal dust is included in our modelling, and cannot be replicated by simply increasing the column of external dust. (For comparison, Fig. \ref{fig:contours_nodust} reproduces the left panels of Fig. \ref{fig:contours} without the inclusion of internal dust.) The effect of the radiation from internal dust on the level populations of the first excited $^2\Pi_{3/2}\,J=5/2$~and the second excited $^2\Pi_{1/2}\,J=1/2$~rotational states (at $\lambda \approx 120\,{\rm \mu m}$~and 80\,$\mu$m, respectively) is illustrated in Fig. \ref{fig:level_ratio}. This figure shows the proportional change in population of the hyperfine levels within these two excited rotational states in the `low' radiation regime, for a gas kinetic temperature of 30\,K, number density of $10^3\,{\rm cm}^{-3}$~and velocity dispersion of 1.4\,\kms. The population in all levels is increased when radiation from internal dust is included, and this increase is greatest for the second excited $^2\Pi_{1/2}\,J=1/2$~rotational state (shown in red in Fig. \ref{fig:level_ratio}), which tends to invert the 1612 MHz line.

The dominant pattern of 1720 MHz inversion seen in the low radiation regime (particularly in the lower gas temperature and number density panels) is qualitatively consistent with the fact that the 1720 MHz line is widely observed in emission in the ambient ISM of the Galactic Plane, in contrast to the 1612 MHz line which is more commonly seen in absorption \citep{Turner1982,Dawson2014a}. Increasing the radiation field increases the range of column densities and velocity dispersions at which the 1612 MHz line will invert, and inhibits the 1720 MHz inversion at low gas temperatures and number densities, while still readily allowing 1720 MHz inversion at higher gas temperatures and number densities. 

In the case of the \flipsperfect~examples of the satellite-line flip where the 1720 MHz inversion is seen at more negative velocities, has a bright background source of continuum, and is associated on sky and in velocity with an \HII~region, we may consider a model in which an \HII~region provides the high radiation field required to invert the 1612 MHz line at typical molecular cloud gas temperatures and number densities. The expanding \HII~region drives a shock into this molecular gas, as illustrated by the cartoon in Fig. \ref{fig:flip_cartoon}. The moderate shock velocity \citep[$<10$\,\kms][]{Hosokawa2005, Hosokawa2006} will only partially dissociate the molecular gas \citep{Draine1983,Flower2003b}. The increased gas temperature and number density of the shock pushes the post-shock gas into the collisionally dominated regime, disabling the 1612 MHz inversion and inverting the 1720 MHz line. This picture provides a very natural explanation for the observed features of the satellite line flip. 

\section{Discussion}\label{Sec:Discussion}
    The high-radiation case outlined above provides a qualitatively convincing explanation of the satellite line flip as a phenomenon driven by the expansion of an \HII~region into a molecular cloud. The on-sky association of the majority of observed flips with known \HII~regions, together with the systematic bias in velocity orientation (with 1720 MHz-inverted gas always approaching faster), paint very consistent parts of this picture. However, it is important to note that a satellite line flip can in principle be produced without the need for expansion, shocks, or a high IR radiation field. The basic requirements are only the presence of molecular gas on either side of a velocity discontinuity, with the gas on either side possessing sufficiently different physical parameters to invert the relevant lines. Our non-LTE modelling indeed shows that a flip could in principle be driven by changes in number density, column density, velocity dispersion or gas temperature, even in ambient low-radiation regions.

That said, of the two radiation fields considered in our modelling (illustrated in Fig. \ref{fig:contours}), a flip is more readily and comfortably produced in the `high' radiation regime. In this scenario, increased FIR radiation allows the 1612 MHz inversion to be radiatively pumped in the pre-shock gas of the molecular cloud -- in contrast to the 1720 MHz emission more generally seen in the ambient molecular ISM \citep{Turner1982,Dawson2014a}. Our modelling has then shown that the increased temperature and number density of the post-shock gas can disable the 1612 MHz pumping mechanism and allow the 1720 MHz line to invert. These are both essential elements of our \HII~region picture, which requires 1612 MHz emission from the ambient pre-shock gas, and 1720 MHz emission from a blue-shifted, shocked component. 

Importantly, however, this high radiation regime also allows the 1612 MHz line to invert at lower column densities, permitting a lower line-of-sight extent for the 1612 MHz-inverted component of the flip. Assuming an ambient molecular cloud density of $n_{\rm H_2}\approx 10^3\,{\rm cm}^{-3}$~and an OH abundance relative to \Htwo~of $X_{\rm OH}=10^{-7}$, the column density of $N_{\rm OH} = 10^{15.5}\,{\rm cm}^{-2}$~shown in the high radiation synthetic spectrum at the bottom right in Fig. \ref{fig:contours}~would require a line-of-sight extent of $\approx 10\,$pc. In contrast, the column density of $N_{\rm OH} = 10^{16}\,{\rm cm}^{-2}$ shown in the low radiation synthetic spectrum would require a line-of-sight extent of $\approx 30$\,pc. Both distances are problematic when we consider the velocity gradient of the ambient molecular cloud that is suggested by these values. The synthetic spectrum in the low radiation regime would require a velocity gradient of $\approx 0.04\,{\rm km\,s}^{-1}\,{\rm pc}^{-1}$, while that of the high radiation regime would require a velocity gradient of $\approx 0.12\,{\rm km\,s}^{-1}\,{\rm pc}^{-1}$: both significantly lower than the expected value of $1\,{\rm km\,s}^{-1}\,{\rm pc}^{-1}$~\citep{Scoville1974}. We could of course make sensible increases to the ambient molecular cloud number density or the OH to \Htwo~abundance to produce a more acceptable velocity gradient, but such fine-tuning of a simplified 1D model may not be reasonable. It is difficult to evaluate this conclusion further without the support of more detailed or complex models. 

\begin{table}
    \centering
    \begin{tabular}{cccccc}
        \hline
         \multicolumn{2}{c}{Flip location}&Velocity&\multicolumn{2}{c}{Background}&Ref$^{c}$\\
         $l^{\circ}$&$b^{\circ}$&orientation$^{a}$&\multicolumn{2}{c}{continuum$^{b}$}&\\
         \hline
         48.92&-0.28&1720 (-)&Extragalactic&41\,K&1\\
         172.80&-13.24&1612 (-)&CMB/synchrotron&4\,K&2\\
         173.40&-13.26&1612 (-)&CMB/synchrotron&4\,K&3\\
         175.83&-9.36&1612 (-)&CMB/synchrotron&4\,K&1\\
         336.95&-0.20&1720 (-)&Diffuse/unresolved&37\,K&4\\
         \hline
    \end{tabular}
    \caption{Approximate 18\,cm background brightness temperatures of the 5 instances of the satellite line flip not associated with \HII~regions. $^{a}$Velocity orientation of the flip is indicated by the line that is inverted at more negative velocities. $^{b}$Description of the background source of continuum and its brightness temperature at $\approx 18$\,cm. $^{c}$References: 1. Petzler et al. (2020 in prep), 2. \citet{Xu2016}, 3. \citet{Ebisawa2019}, 4. \citet{Dawson2014a}.}
    \label{tab:weird_flips}
\end{table}

In our sample there are five instances of the flip without associated RRL detections. Table \ref{tab:weird_flips}~lists their background brightness temperatures and restates their velocity orientations. First, we consider the flips with the more common velocity orientation (1720 MHz line inverted at more negative velocities), which are seen towards G048.92-0.28 (Petzler et al. 2020 in prep) and G336.95-0.20 \citep[$-40\,$\kms,][]{Dawson2014a}. Both of these sources lie in the Galactic Plane, have high background continuum temperatures and are positionally coincident with regions of active star formation. While they do not have literature RRL detections within 10\,\kms~of either component of the flip, it seems likely that these examples may represent pre- and post-shock gas surrounding as-yet unidentified \HII~regions.

Next, we consider the flips with the opposite velocity orientation (1612 MHz line inverted at more negative velocities), which are seen towards G172.80-13.24 \citep{Xu2016}, G173.40-13.26 \citep{Ebisawa2019} and G175.83-9.36 (Petzler et al. 2020 in prep). All three of these examples are within the Taurus molecular cloud complex and have low background continuum brightness temperature ($T_{\rm c}\sim 4$\,K). The flip towards G175.83-9.36 is seen in optical depth with the 1612 MHz emission at $\tau = -0.003$~and the 1720 MHz at $\tau = -0.03$ (Petzler et al. 2020 in prep). Our modelling indicates that such weak inversion of the 1612 MHz line is possible in the low radiation regime at quite low column density. The 1720 MHz inversion could also be readily produced in this radiation regime across a wide range of column densities. Our modelling does not suggest that this example of the flip implies the presence of an enhanced radiation field or a shock. 

The other two examples in Taurus (G172.80-13.24 and G173.40-13.26) are both reported in brightness temperature only, and it is therefore possible that they are not in fact stimulated emission ($T_{\rm ex}<0$), but rather conventional thermal emission ($T_{\rm ex}>T_{\rm c}$). The pumping mechanisms outlined in Section \ref{Sec:Excitation} could still be responsible for the deviations from LTE in this gas, but such thermal emission would not correspond exactly to the regions of stimulated emission illustrated in Fig. \ref{fig:contours}. \citet{Ebisawa2019}~have recently presented non-LTE excitation modelling similar to ours of the flip towards G173.40-13.26. Their modelling focused on the blue-shifted component of this flip. They assumed a `typical' velocity dispersion of 1\,\kms, and a dust temperature of 15\,K based on IR observations \citep{Flagey2009}. They found that the sub-thermal excitation of the 1720 MHz line inferred from their observations was best fit by a model with extremely high external dust column density (\Av~$ \approx 150$\,mag), high OH column density ($N_{\rm OH} > 10^{15.5}\,{\rm cm}^{-2}$) and low gas temperature ($T_{\rm gas} < 30\,$K). However, they did not propose a physical mechanism by which such a high dust column could exist in Taurus. 

A key difference between our modelling and that of \citet{Ebisawa2019} appears to be our inclusion of the effects of internal dust: when we remove this from our modelling, the weak 1612 MHz inversion seen at low column densities in the low radiation regime in Fig. \ref{fig:contours} is not present. This weak 1612 MHz inversion is simultaneous with sub-thermal excitation of the 1720 MHz line, implying that the inclusion of internal dust may allow the observations of \citet{Ebisawa2019} to be modelled without the need for such a high external dust column. Indeed, we find that our modelling is able to approximately reproduce the observed brightness temperatures of their blue-shifted component (with the same beam-filling factor used by \citet{Ebisawa2019} of 0.1) in a low radiation environment.

The above discussion highlights the need for caution when considering the exact quantitative results of 1D molecular excitation modelling. There are several known factors that our modelling does not consider, such as the effects of turbulence, unresolved structure in the foreground gas, and uncertainties or errors in collisional rates. Indeed, new collision rates of OH and \Htwo~\citep{Klos2020} are systematically higher than those used of \cite{Offer1994} in our modelling. There are also other significant observational assumptions that we do not address, such as structure in the background continuum source \citep{Engelke2019}. Therefore the fine-tuning of our model parameters in order to match observations is not likely to be strictly valid. Nevertheless, the qualitative physical picture presented here is convincing: the satellite line flip is most commonly and readily produced in pre- and post-shock gas in the high-radiation environment surrounding an expanding \HII~region. 

\section{Conclusions}\label{Sec:Conclusions}
    In this work we present \totalflips~examples from the literature of the OH satellite-line flip: a peculiar profile wherein both of the OH ground rotational state satellite lines flip -- one from emission to absorption and the other the reverse -- across a closely blended double feature. We note that with the exception of the three flips observed towards the Taurus molecular cloud, all are observed towards bright background continuum and have the same orientation in velocity: the 1720 MHz inversion is seen at more negative velocities and the 1612 MHz inversion at more positive velocities. In addition, we note that all save \flipsnoHII~are associated with radio recombination lines of \HII~regions.

We find that the 1612 MHz inversion can be radiatively pumped in an environment with typical molecular cloud gas temperature of $T_{\rm gas} \approx 20$ -- $30\,$K and number density of $n_{\rm H_2} \approx 10^2$ -- $10^3\,{\rm cm}^{-3}$, that is exposed to a high radiation field. The 1720 MHz inversion can then originate in gas exposed to this same radiation field but with enhanced gas temperature ($T_{\rm gas} \approx 50$ -- $100\,$K) and number density ($n_{\rm H_2} \approx 10^4$ -- $10^5\,{\rm cm}^{-3}$). We demonstrate that these physical conditions can be realised by a shock expanding from a central \HII~region: the \HII~region provides the enhanced radiation field required to invert the 1612 MHz line in the pre-shock gas while the shock heats and compresses the gas sufficiently to invert the 1720 MHz line. The brightness of the central \HII~region causes the foreground gas to dominate the observed brightness temperature spectra, so the 1720 MHz inversion in the post-shock gas is blue-shifted relative to the 1612 MHz inversion in the pre-shock gas. We support this model through our own non-LTE molecular excitation modelling, and with reference to observations \citep{Okumura1996,Anderson2012,Deharveng2012} and modelling \citep{Churchwell1990,Hosokawa2005,Hosokawa2006} of \HII~regions and dust temperatures. The precise quantitative results of our simplified 1D modelling should be regarded with caution as our models do not account for the clumpy and filamentary structure, nor the effects of turbulent motions known to exist in molecular clouds. The inclusion of such mechanisms may allow us to more precisely match the observed OH spectra.

The results presented in this work have highlighted a potential new tool for pinpointing the locations and properties of molecular gas associated with \HII~regions. In principle, observations of the OH satellite-line flip should allow us to characterise the physical conditions in the molecular gas interacting with numerous \HII~regions throughout the Galactic Disk, providing information on kinematics, density and temperature. For example, the velocity spacing of the flip and modelling of the blue-shifted 1720 MHz inversion could then characterise the kinematics and physical conditions of the expanding shock. As a radio spectral line probe, the OH lines (like RRLs) allow us to see through the full depth of the Milky Way disk, and localise interaction sites at least in the spatio-velocity domain, if not in true 3D space as well. Cross-matching with multi-wavelength tracers of \HII~regions in the Milky Way Disk would therefore allow us to paint a more complete picture of the interactions between \HII~regions and molecular gas in the Milky Way.  
With a number of large-scale OH Galactic Plane surveys either published or underway (e.g. THOR; \citealt{Beuther2016}, SPLASH; \citealt{Dawson2014a}, and GASKAP; \citealt{Dickey2013}), there is scope to identify far more examples of the satellite flip, and moreover to model the overall trends in all four transitions throughout large regions of the Milky Way.

\section*{Data Availability}
The data underlying this article were obtained from \citet{Goss1967}, 
\citet{Manchester1970}, 
\citet{Caswell1974}, 
\citet{Turner1979}, 
\citet{vanLangevelde1995}, 
\citet{Frayer1998}, 
\citet{Brooks2001},
\citet{Dawson2014a}, 
\citet{Xu2016}, 
\citet{Rugel2018}, 
\citet{Ebisawa2019} and 
\citet{Ogbodo2020}, as outlined in Table \ref{Tab:Flip}. Additional data from Petzler et al. (2020 in prep.) and from private communication will be shared on reasonable request to the corresponding author.
\section*{Acknowledgements}

This work was inspired in part by discussions of the flip held at any and all opportunities over the span of several years. Thus we acknowledge the contributions and interest of Maria Cunningham, John Dickey, Carl Heiles, Ron Allen, Philip Engelke, Claire Murray, Shu-ichiro Inutsuka, Satoshi Yamamoto and others. We sincerely thank the anonymous reviewer for their detailed and thoughtful feedback which greatly improved this work. 
A.P. is the recipient of an Australian Government Research Training Program (RTP) stipend and tuition fee offset scholarship. J.R.D. is the recipient of an Australian Research Council (ARC) DECRA Fellowship (project number DE170101086), which partially supported this work. 

\bibliographystyle{mnras}
\bibliography{Bibliography}

\appendix
\section{Satellite-line inversion without internal dust}
    \begin{figure*}
    \centering
    \begin{tabular}{cc}
    
    \thead[cc]{`Low' Radiation:\\
    $T_{\rm dust}({\rm ext})=20$\,K\\
    \Av~$=0.3$\,mag\\
    \\
    \includegraphics[trim = {5.7cm 1.4cm 1.3cm 8.9cm}, clip = true, width=0.15\linewidth]{Contour_plot_legend.pdf}}&
    \thead[cc]{\includegraphics[trim = {-.6cm 0cm -.6cm .5cm}, clip = true, width=0.42\linewidth]{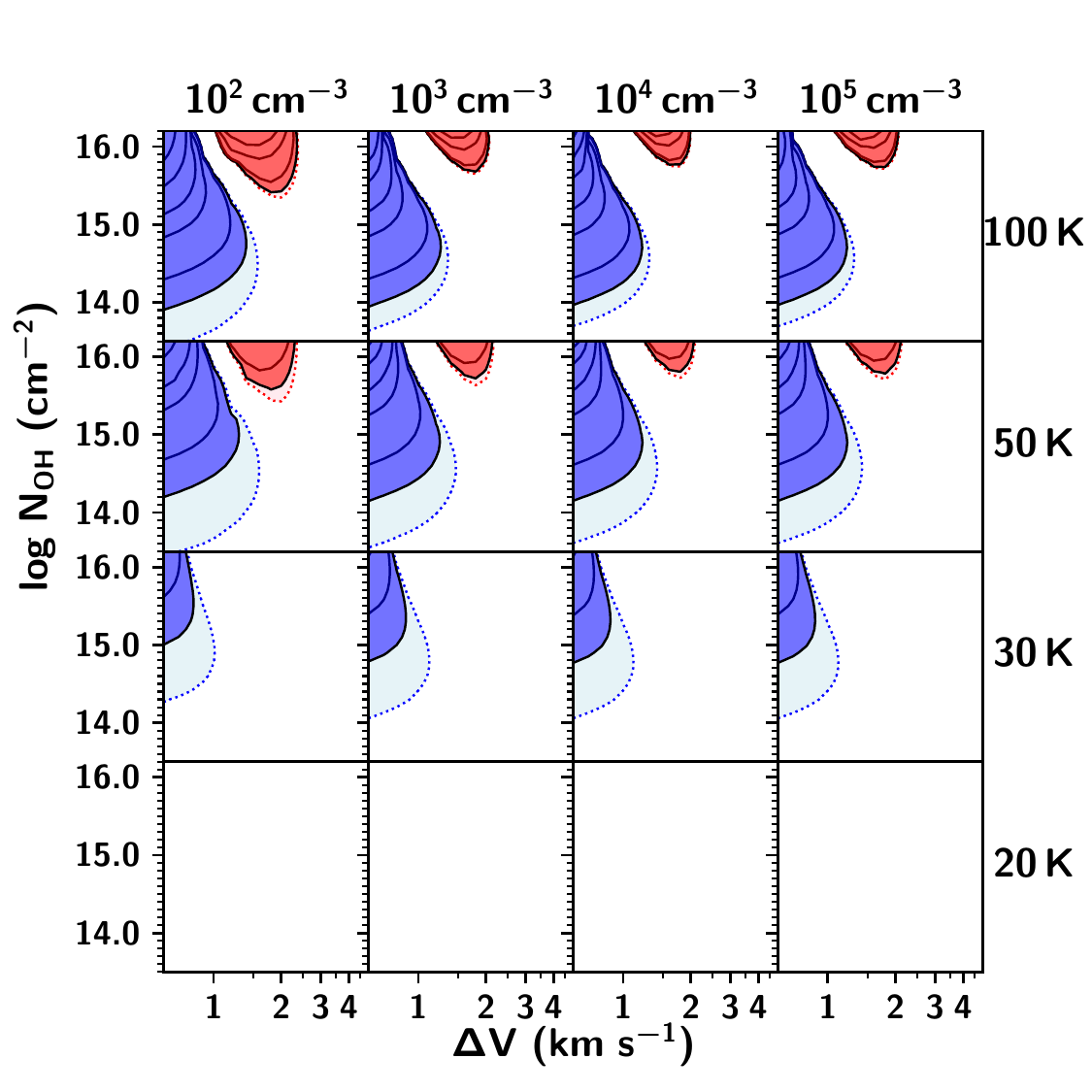}}\\
    
    \thead[cc]{`High' Radiation:\\
    $T_{\rm dust}({\rm ext})=70$\,K\\
    \Av~$=0.1$\,mag\\
    \\
    \includegraphics[trim = {5.7cm 1.4cm 1.3cm 8.9cm}, clip = true, width=0.15\linewidth]{Contour_plot_legend.pdf}}&
    \thead[cc]{\includegraphics[trim = {-.6cm 0cm -.6cm .5cm}, clip = true, width=0.42\linewidth]{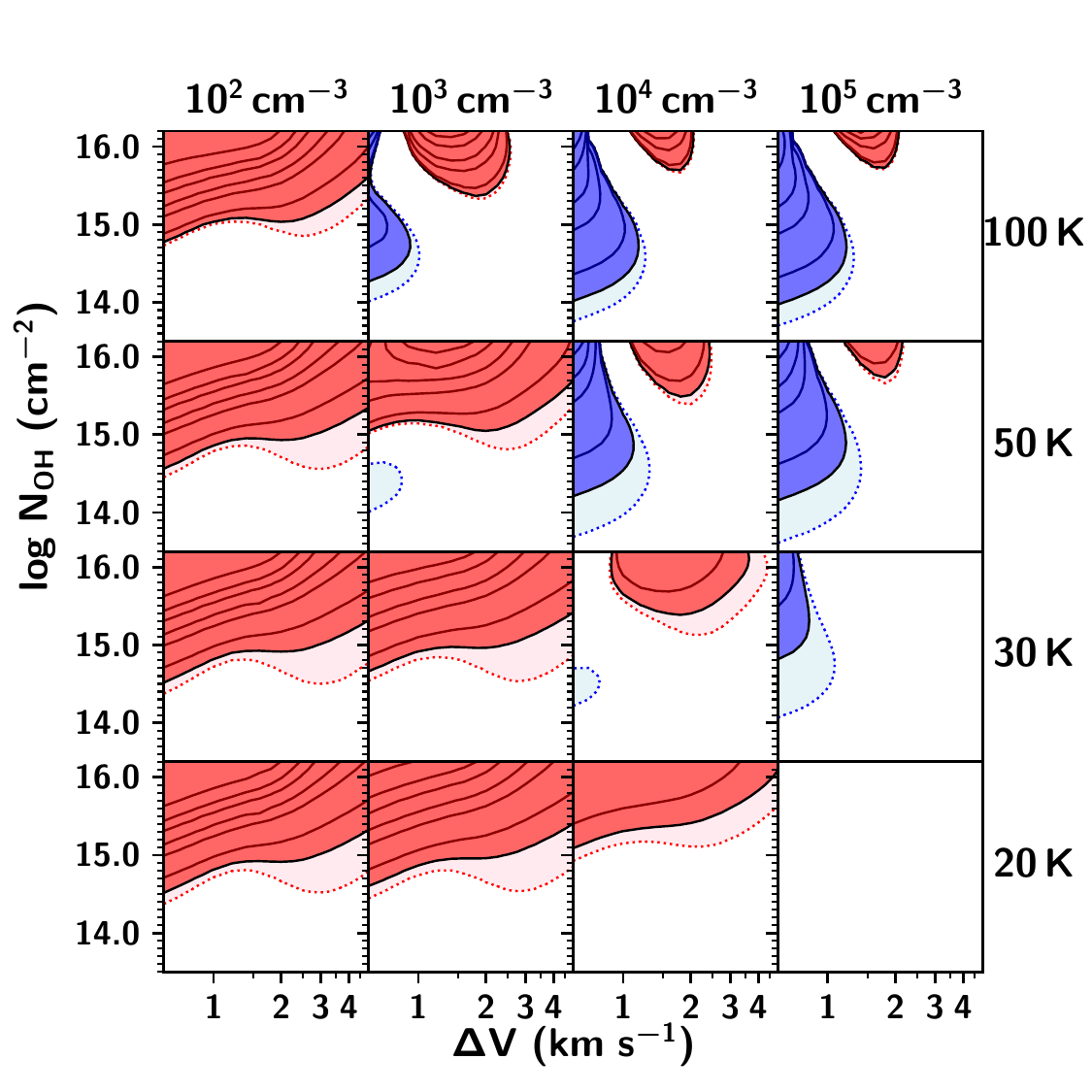}}\\
    
    \end{tabular}
    \caption{Same as the left panels of Fig. \ref{fig:contours} without the effects of the radiation from internal dust.}
    \label{fig:contours_nodust}
\end{figure*}

\bsp	
\label{lastpage}
\end{document}